\documentclass{moriond}

\usepackage{amsmath} 
\usepackage{amssymb}
\usepackage{amsfonts}
\usepackage{ifthen}
\newboolean{inbibliography}
\setboolean{inbibliography}{false} 
\newboolean{uprightparticles}
\setboolean{uprightparticles}{false} 
\usepackage{xspace}
\usepackage{upgreek}
\usepackage{hyperxmp}
\usepackage{hyperref}
\usepackage[superscript,nomove]{cite} 
\usepackage{mciteplus}
\newboolean{articletitles}
\setboolean{articletitles}{false} 

\def\CP{{\ensuremath{C\!P}}\xspace}
\newcommand{\decay}[2]{\mbox{\ensuremath{#1\!\to #2}}\xspace}
\def\PK      {\ensuremath{K}\xspace}
\def\kaon    {{\ensuremath{\PK}}\xspace}
\def\Kp      {{\ensuremath{\kaon^+}}\xspace}
\def\Km      {{\ensuremath{\kaon^-}}\xspace}
\def\Ppi         {\ensuremath{\pi}\xspace}
\def\pion   {{\ensuremath{\Ppi}}\xspace}
\def\pip    {{\ensuremath{\pion^+}}\xspace}
\def\pim    {{\ensuremath{\pion^-}}\xspace}
\def\Dz      {{\ensuremath{\D^0}}\xspace}
\def\Dbar    {{\kern 0.2em\overline{\kern -0.2em \PD}{}}\xspace}
\def\Dzb     {{\ensuremath{\Dbar{}^0}}\xspace}
\def\dkk        {\decay{\Dz}{\Km\Kp}}
\def\dpipi      {\decay{\Dz}{\pim\pip}}

\newcommand{\aunit}[1]{\ensuremath{\text{\,#1}}}

\def\fb   {\ensuremath{\aunit{fb}}\xspace}
\def\invfb   {\ensuremath{\fb^{-1}}\xspace}
\newcommand{\tev}{\aunit{Te\kern -0.1em V}\xspace}
\def\PD      {\ensuremath{D}\xspace}
\def\D       {{\ensuremath{\PD}}\xspace}
\def\Dstar   {{\ensuremath{\D^*}}\xspace}
\def\PB      {\ensuremath{B}\xspace}
\def\B       {{\ensuremath{\PB}}\xspace}
\def\Bbar    {{\ensuremath{\kern 0.18em\overline{\kern -0.18em \PB}{}}}\xspace}
\def\Bb      {{\ensuremath{\Bbar}}\xspace}
\newcommand{\deltaACP}{\ensuremath{\Delta A_{\CP}}\xspace}
\newcommand{\DACP}{\deltaACP}
\def\Pp      {\ensuremath{p}\xspace}
\def\proton      {{\ensuremath{\Pp}}\xspace}
\def\Dstarp  {{\ensuremath{\D^{*+}}}\xspace}
\def\Dstarm  {{\ensuremath{\D^{*-}}}\xspace}
\newcommand{\Araw}{\ensuremath{A_{\rm raw}}\xspace}
\newcommand{\AD}{\ensuremath{A_{\rm D}}\xspace}
\newcommand{\AP}{\ensuremath{A_{\rm P}}\xspace}
\def\order   {{\ensuremath{\mathcal{O}}}\xspace}
\def\Pb      {\ensuremath{b}\xspace}
\def\bquark    {{\ensuremath{\Pb}}\xspace}
\def\Pc      {\ensuremath{c}\xspace}
\def\cquark    {{\ensuremath{\Pc}}\xspace}
\def\cquarkbar {{\ensuremath{\overline \cquark}}\xspace}
\newcommand{\chisqip}{\ensuremath{\chi^2_{\text{IP}}}\xspace}

\newcommand{\mevcc}{\ensuremath{\aunit{Me\kern -0.1em V\!/}c^2}\xspace}

\def\dkpi       {\decay{\Dz}{\Km\pip}}

\def\piz    {{\ensuremath{\pion^0}}\xspace}
\def\Pmu         {\ensuremath{\mu}\xspace}
\def\mup        {{\ensuremath{\Pmu^+}}\xspace}
\def\Pnu         {\ensuremath{\nu}\xspace}
\def\neu        {{\ensuremath{\Pnu}}\xspace}
\def\neum       {{\ensuremath{\neu_\mu}}\xspace}
\def\Pe      {\ensuremath{e}\xspace}
\def\ep         {{\ensuremath{\Pe^+}}\xspace}
\def\neue       {{\ensuremath{\neu_e}}\xspace}
\newcommand{\APB}{\ensuremath{A_{\rm P}(\PB)}\xspace}

\newcommand{\mean}[1]{\ensuremath{\left\langle #1 \right\rangle}}
\def\agamma     {\ensuremath{A_{\Gamma}}\xspace}

\def\PLambda     {\ensuremath{\Lambda}\xspace}
\mathchardef\PLambda="7103

\newcommand{\Photo}{}

\begin{document}
\vspace*{4cm}
\title{OBSERVATION OF \boldmath{\CP} VIOLATION IN CHARM DECAYS AT LHCb}

\author{ F. BETTI on behalf of the LHCb collaboration}

\address{Universit\`a di Bologna, Dipartimento di Fisica e Astronomia,\\
Istituto Nazionale di Fisica Nucleare - Sezione di Bologna,\\
viale Berti Pichat 6/2, Bologna (40127), Italy}

\maketitle\abstracts{
A search for charge-parity~($C\!P$) violation in $D^0 \to K^- K^+$ and $D^0 \to \pi^-\pi^+$ decays is reported, using $pp$ collision data corresponding to an integrated luminosity of 5.9~$\mathrm{fb}^{-1}$ collected at a center-of-mass energy of 13~TeV with the LHCb detector. The flavor of the $D^0$ meson is determined from the charge of the pion in $D^*(2010)^+ \to D^0 \pi^+$ decays or from the charge of the muon in $\kern 0.18em\overline{\kern -0.18em B}{} \to D^0 \mu^- \bar{\nu}_\mu X$ decays. The difference between the $C\!P$ asymmetries in $D^0 \to K^- K^+$ and $D^0 \to \pi^-\pi^+$ decays is measured to be \mbox{$\Delta A_{C\!P} = [ -18.2 \pm 3.2\,(\rm stat.) \pm 0.9\,(\rm syst.) ] \times 10^{-4}$} for $\pi$-tagged and \mbox{$\Delta A_{C\!P} = [ -9 \pm 8\,(\rm stat.) \pm 5\,(\rm syst.) ] \times 10^{-4} $} for $\mu$-tagged $D^0$ mesons. The combination with previous LHCb results leads to $$\Delta A_{C\!P} = ( -15.4 \pm 2.9) \times 10^{-4},$$ where the uncertainty includes both statistical and systematic contributions. The measured value differs from zero by more than five standard deviations. This is the first observation of \CP violation in the decay of charm hadrons.
}

\section{Introduction}

The noninvariance of fundamental interactions under the combined action of charge conjugation~($C$) and parity~($P$) transformations, so-called \CP violation, is a necessary condition for the dynamical generation of the baryon asymmetry of the universe~\cite{Sakharov:1967dj}.
\CP violation is included in the Standard Model~(SM) of particle physics through an irreducible complex phase in the Cabibbo-Kobayashi-Maskawa~(CKM) quark-mixing matrix~\cite{Cabibbo:1963yz,Kobayashi:1973fv}.
Several experiments established the presence of \CP violation in weak interactions in the $K$- and $B$-meson systems~\cite{Christenson:1964fg,AlaviHarati:1999xp,Lai:2001ki,Aubert:2001nu,Abe:2001xe,Aubert:2004qm,Chao:2004mn,LHCb-PAPER-2013-018,LHCb-PAPER-2012-001}, and all results are well interpreted within the CKM formalism.
However, the size of \CP violation in the SM is too small to account for the observed matter-antimatter asymmetry~\cite{Cohen:1993nk,Riotto:1999yt,Hou:2008xd}, suggesting the existence of beyond-the-SM sources of \CP violation.

Despite decades of experimental searches, the observation of \CP violation in the charm sector has not yet been achieved.
Because of the presence of low-energy strong-interaction effects, theoretical predictions  of the size of \CP violation in charm decays are difficult to compute reliably, and the asymmetries are expected to be of the order of $10^{-4}$--$10^{-3}$ in magnitude~\cite{Golden:1989qx,Buccella:1994nf,Bianco:2003vb,Grossman:2006jg,Artuso:2008vf,Brod:2011re,Cheng:2012wr,Cheng:2012xb,Li:2012cfa,Franco:2012ck,Pirtskhalava:2011va,Feldmann:2012js,Brod:2012ud,Hiller:2012xm,Grossman:2012ry,Bhattacharya:2012ah,Muller:2015rna,Khodjamirian:2017zdu,Buccella:2019kpn}. 

Searches for \CP violation in  \dkk and \dpipi modes\footnote{The inclusion of charge-conjugate decay modes is implied throughout except in asymmetry definitions.} have been performed by the BaBar~\cite{bib:babarpaper2008}, Belle~\cite{bib:bellepaper2008}, CDF~\cite{bib:cdfpaper,CDF:2012qw} and LHCb~\cite{LHCB-PAPER-2011-023,LHCB-PAPER-2013-003,LHCB-PAPER-2014-013,LHCB-PAPER-2015-055,LHCB-PAPER-2016-035} collaborations, which measured values of \CP asymmetries consistent with zero within a precision of a few per mille.
This document presents a measurement of the difference of the time-integrated \CP asymmetries in \dkk and \dpipi decays, performed using \proton\proton collision data collected with the LHCb detector between 2015 and 2018 at a center-of-mass energy of 13\tev, corresponding to an integrated luminosity of 5.9\invfb. 

The time-dependent \CP asymmetry, $A_{\CP}(f;\,t)$, between states produced as \Dz or \Dzb mesons decaying to a \CP eigenstate $f$ at time $t$ is defined as
\begin{equation}
 \label{eq:acpf}
 	A_{\CP}(f;\,t) \equiv \frac{\Gamma(\Dz(t) \to f)-\Gamma(\Dzb(t) \to f)}{\Gamma(\Dz(t) \to f)+\Gamma(\Dzb(t) \to f)},
\end{equation}
where $\Gamma$ denotes the time-dependent rate of a given decay.
For $f= \Km\Kp$ or $f= \pim\pip$, $A_{\CP}(f;\,t)$ can be expressed in terms of a direct component associated to \CP violation in the decay amplitude and another component associated to \CP violation in \Dz--\Dzb mixing or in the interference between mixing and decay. 
The corresponding time-integrated asymmetry, $A_{\CP}(f)$, can be written to first order in the \Dz--\Dzb mixing parameters as~\cite{bib:cdfpaper,Gersabeck:2011xj}
\begin{equation}
\label{eq:acpphysics}
	A_{\CP}(f)  \approx a_{\CP}^{\rm dir}(f) - \frac{\langle t (f)\rangle}{\tau(\Dz)}\,A_\Gamma(f),
\end{equation}
where $\langle t (f)\rangle$ denotes the mean decay time of $\Dz\to f$ decays in the reconstructed sample, $a_{\CP}^{\rm dir}(f)$ is the direct \CP asymmetry, $\tau(\Dz)$ the \Dz lifetime and $A_\Gamma(f)$ the asymmetry between the $\Dz \to f$ and $\Dzb \to f$ effective decay widths~\cite{LHCb-PAPER-2014-069,LHCb-PAPER-2016-063}.
Taking $A_\Gamma$ to be independent of the final state~\cite{Grossman:2006jg,bib:kagansokoloff,Du:2006jc}, the difference between \CP asymmetries in \dkk and \dpipi decays is
\begin{eqnarray}
\label{eq:dacpdef1}
	\DACP  & \equiv &  A_{\CP}(\Km\Kp) - A_{\CP}(\pim\pip) \nonumber \\
	& \approx & \Delta a_{\CP}^{\rm dir} - \frac{\Delta \langle t \rangle}{\tau(\Dz)}\,A_\Gamma,
\label{eq:dacpdef}
\end{eqnarray}
where $\Delta a_{\CP}^{\rm dir} \equiv a_{\CP}^{\rm dir} (K^-K^+) - a_{\CP}^{\rm dir} (\pi^-\pi^+)$ and $\Delta \langle t \rangle$ is the difference of the mean decay times $\langle t (\Km\Kp)\rangle$ and $\langle t (\pim\pip)\rangle$.

The \Dz mesons considered in this analysis are produced in two ways: promptly at a \proton\proton collision point (primary vertex, PV) in the strong \decay{\Dstar(2010)^+}{ \Dz \pip}decay~(hereafter $\Dstar(2010)^+$ is referred to as $\Dstarp$) or at a vertex displaced from any PV in semileptonic \decay{\Bb}{\Dz \mu^- \bar{\nu}_\mu X} decays, where $\Bb$ denotes a hadron containing a $b$ quark and $X$ stands for additional particles.
The flavor at production of \Dz mesons from $\Dstarp$ decays is obtained from the charge of the accompanying pion~($\pi$-tagged), whereas that of \Dz mesons from semileptonic $b$-hadron decays is determined from the charge of the accompanying muon~($\mu$-tagged).
The raw asymmetries measured for $\pi$-tagged and $\mu$-tagged \Dz decays are defined as
\begin{align}
	\begin{split}
		\Araw^{\pi{\mbox{-}}\rm{tagged}}(f)& \equiv \frac {N\left(\Dstarp \to \Dz(f)\pi^{+}\right) - N\left(\Dstarm \to \Dzb(f)\pi^{-}\right)} {N\left(\Dstarp \to \Dz(f)\pi^{+}\right) + N\left(\Dstarm \to \Dzb(f)\pi^{-}\right)}, \\
		\Araw^{\mu{\mbox{-}}\rm{tagged}}(f)& \equiv \frac {N(\Bb  \to \Dz(f)\, \mu^- \bar{\nu}_\mu X) - N(\B \to \Dzb(f)\, \mu^+ \nu_\mu X)} { N(\Bb \to \Dz(f)\, \mu^- \bar{\nu}_\mu X) + N(\B \to \Dzb(f)\, \mu^+ \nu_\mu X)},
	\end{split}
\label{eq:araw}
\end{align}
where $N$ is the measured signal yield for each given decay. These can be approximated as
\begin{align}
	\begin{split}
		\Araw^{\pi{\mbox{-}}\rm{tagged}}(f)& \approx A_{\CP}(f) + \AD(\pi) + \AP(\Dstar),\\
		\Araw^{\mu{\mbox{-}}\rm{tagged}}(f)& \approx A_{\CP}(f) + \AD(\mu) + \AP(\B),
	\end{split}
\label{eq:arawstarcomponents}
\end{align}
where $\AD(\pi)$ and $\AD(\mu)$ are detection asymmetries due to different reconstruction efficiencies between positive and negative tagging particles, whereas $\AP(\Dstar)$ and $\AP(\B)$ are the production asymmetries of \Dstar mesons and $b$ hadrons, arising from the hadronization of charm and beauty quarks in \proton\proton collisions.
The involved terms, averaged over phase space for selected events are $\order(10^{-2})$ or less~\cite{LHCb-PAPER-2016-062,LHCb-PAPER-2013-033,LHCb-PAPER-2012-026,LHCb-PAPER-2012-009}, hence the approximations in Eqs.~\ref{eq:arawstarcomponents} are valid up to corrections of $\order(10^{-6})$.
The values of the detection and production asymmetries are independent of the final state $f$, and thus cancel in the difference, if the kinematic distributions of the two channels are equal, resulting in
\begin{equation}
\label{DACP1}
	\DACP = \Araw(\Km\Kp) - \Araw(\pim\pip).
\end{equation}
The relation between $\DACP$ and the measurable raw asymmetries in $\Km\Kp$ and $\pim\pip$ makes the determination of $\DACP$ largely insensitive to systematic uncertainties.

\section{Selection}

The LHCb detector is a single-arm forward spectrometer designed for the study of particles containing \bquark or \cquark quarks~\cite{Alves:2008zz,LHCb-DP-2014-002}.
The magnetic-field polarity of the dipole magnet used by the LHCb tracking system is reversed periodically during data taking to mitigate the differences of reconstruction efficiencies of particles with opposite charges, though the analysis presented in this document is expected to be insensitive to such effects.

The online event selection is performed by a trigger, which consists of a hardware stage based on information from the calorimeter and muon systems, followed by two software stages.
\Dz candidates are fully reconstructed in the second software stage using kinematic, topological and particle-identification~(PID) criteria.
In the $\mu$-tagged sample, \Dz candidates are combined with muons to form \B candidates, under the requirement that they are consistent with originating from a common vertex.
In addition, requirements on the invariant mass of the $\Dz\mu$ system, $m(\Dz\mu)$, and on the corrected mass\footnote{The corrected mass is defined as $m_{\rm corr} \equiv \sqrt{m(\Dz\mu)^2 + p_\perp(\Dz\mu)^2} + p_\perp(\Dz\mu)$~\cite{Kodama:1991ij}, where $p_\perp(\Dz\mu)$ is the momentum of the $\Dz\mu$ system transverse to the flight direction of the $b$ hadron.} are applied in the $\mu$-tagged sample.

In certain kinematic regions very large raw asymmetries, up to 100\%, occur because, for a given magnet polarity, low-momentum particles of one charge at small or large polar angles in the horizontal plane may be deflected out of the detector or into the LHC beam pipe, whereas particles with the other charge are more likely to remain within the acceptance.
For this reason, in the offline selection, fiducial requirements are imposed to exclude kinematic regions characterized by large detection asymmetries for the tagging particle.
About 35\% and 10\% of the selected candidates are rejected by these fiducial requirements for the $\pi$-tagged and $\mu$-tagged samples, respectively.
For $\pi$-tagged \Dz mesons, a requirement on the \Dz \chisqip is applied\footnote{The \chisqip is defined as the difference between the $\chi^2$ of the PV reconstructed with and without the considered particle.} to suppress the background of \Dz mesons produced in $B$ decays, and PID requirements on the \Dz decay products are tightened.
The \Dz and pion candidates are combined to form \Dstarp candidates by requiring a good fit quality of the \Dstarp vertex, that is constrained to coincide with the nearest PV~\cite{Hulsbergen2005566}.
The invariant mass of \Dz candidates is required to lie within a range of about $\pm3$ standard deviations around the known $\Dz$ mass.
For $\mu$-tagged mesons, in order to suppress the combinatorial background due to random combinations of charged kaon or pion pairs not originating from a \Dz decay, the \B candidates are further filtered using a dedicated boosted decision tree~(BDT) that uses variables related to the topology and the kinematics of the reconstructed decay.
A veto in the invariant mass of the $\mu^\mp \pi^\pm$~($\mu^\mp K^\pm$) pair, where the pion~(kaon) is given the muon mass hypothesis, is applied to suppress background from $b$-hadron decays to $\cquark\cquarkbar \pi^\pm X$ ($\cquark\cquarkbar K^\pm X$), where the $\cquark\cquarkbar$ resonance decays to a pair of muons.

The data sample includes events with multiple \Dstarp and \B candidates, that are mostly due to a common reconstructed \Dz meson combined with different tagging particles.
The fractions of events with multiple candidates are about 10\% and 0.4\% in the $\pi$-tagged and $\mu$-tagged samples, respectively.
When multiple candidates are present in the event, only one is kept randomly.

Since the detection and production asymmetries are expected to depend on the kinematics of the reconstructed particles, the possible difference between the kinematic distributions of reconstructed \Dstarp or \B candidates and of the tagging pions or muons in the $K^-K^+$ and $\pi^-\pi^+$ decay modes may induce an incomplete cancellation in the difference in Eq.~\ref{DACP1}.
Hence, a small correction to the $K^-K^+$ sample is applied by means of a weighting procedure: for the $\pi$-tagged sample, the ratio between the three-dimensional background-subtracted distributions of pseudorapidity, transverse momentum and azimuthal angle of the \Dstarp meson in the $K^-K^+$ and $\pi^-\pi^+$ modes is taken and candidate-by-candidate weights are calculated.
An analogous procedure is followed for the $\mu$-tagged sample, where \Dz distributions are used in place of those of the \Dstarp meson.
It is then checked \emph{a posteriori} that the distributions of the same variables for tagging pions and muons are also equalized by the weighting.
The application of the weights results in a small variation of \DACP, below $10^{-4}$ for both the $\pi$-tagged and $\mu$-tagged samples.   

\begin{figure}[t]
\centering
\includegraphics[width=0.45\textwidth]{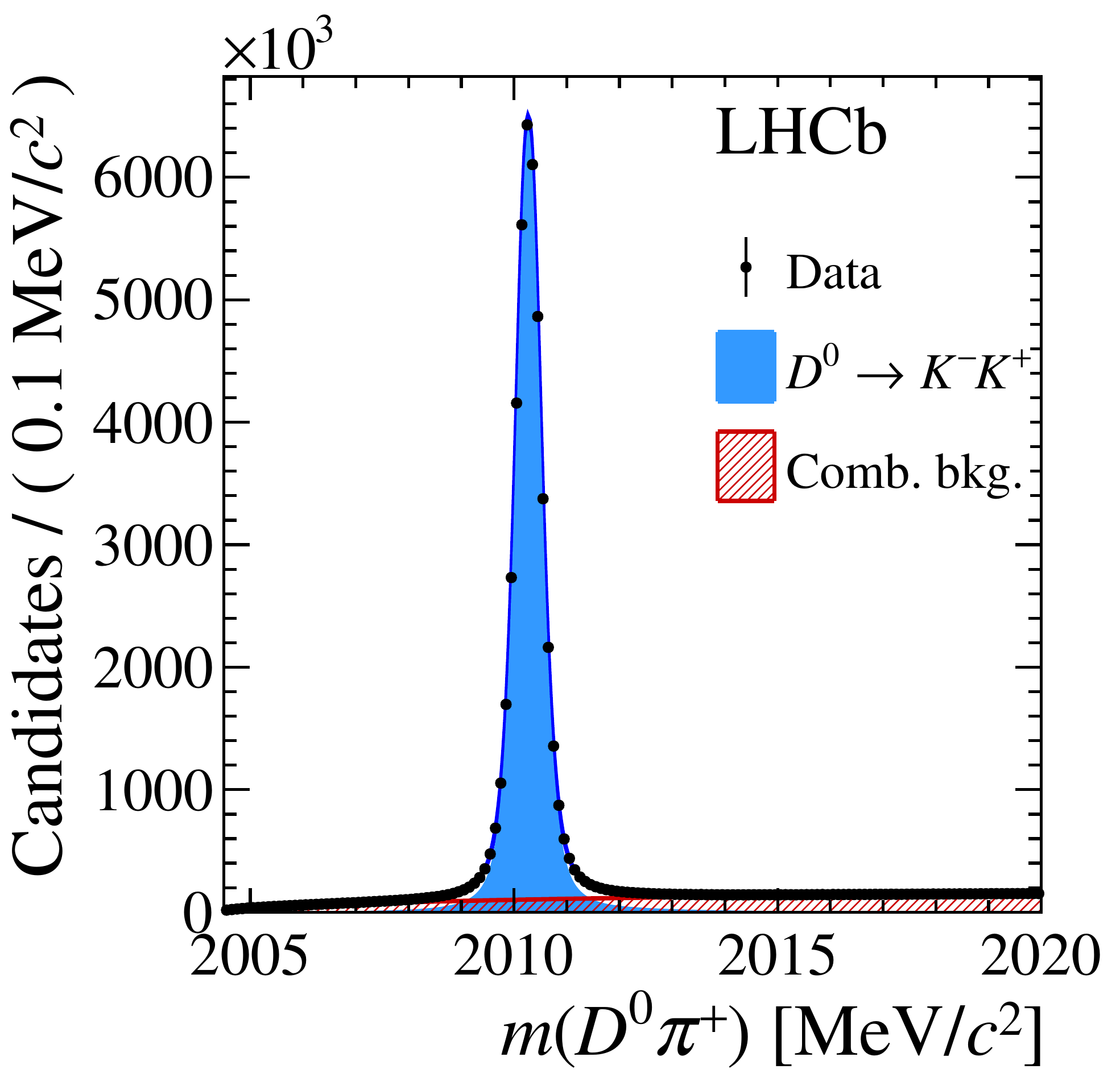}
\includegraphics[width=0.45\textwidth]{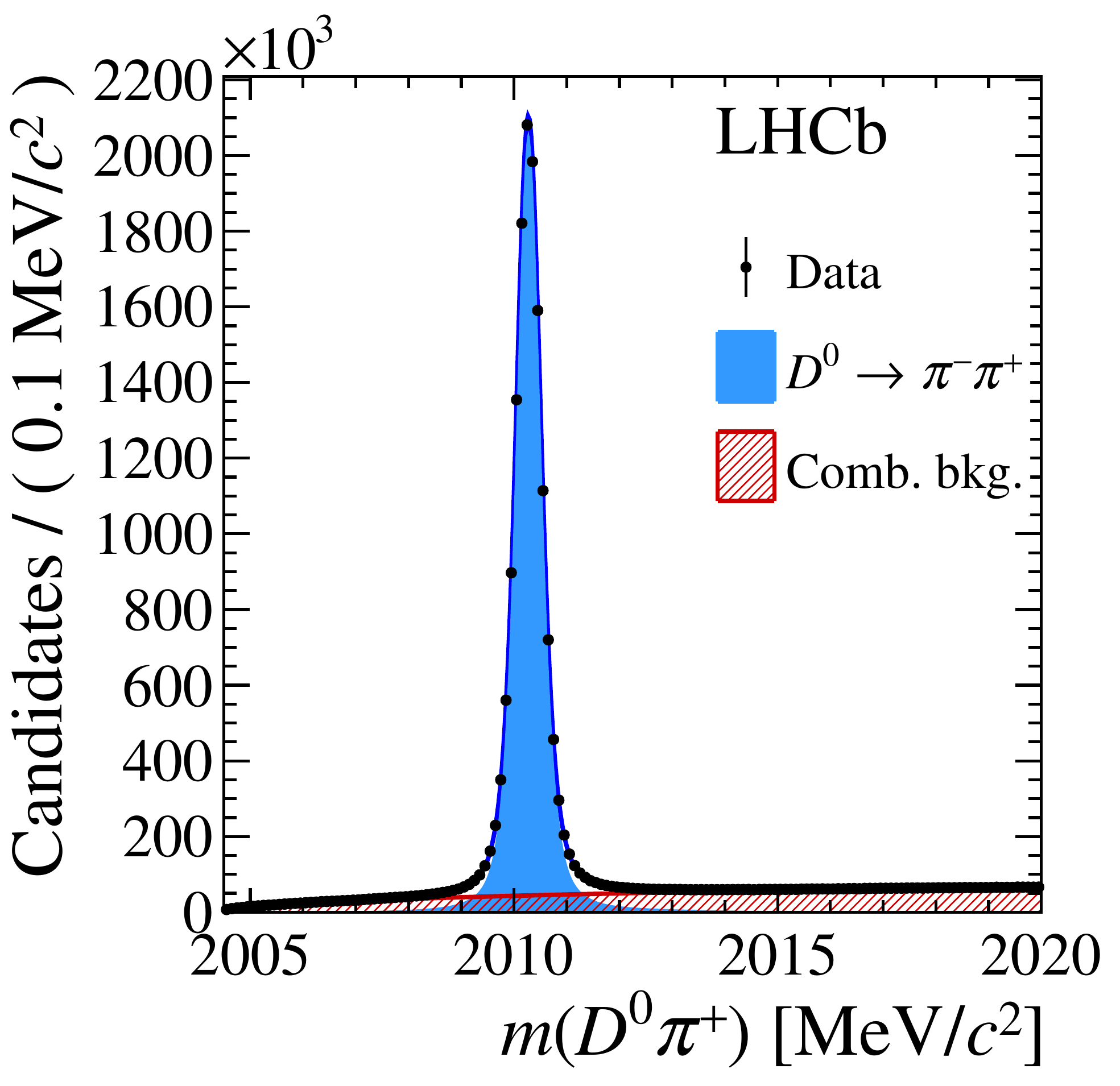}
\includegraphics[width=0.45\textwidth]{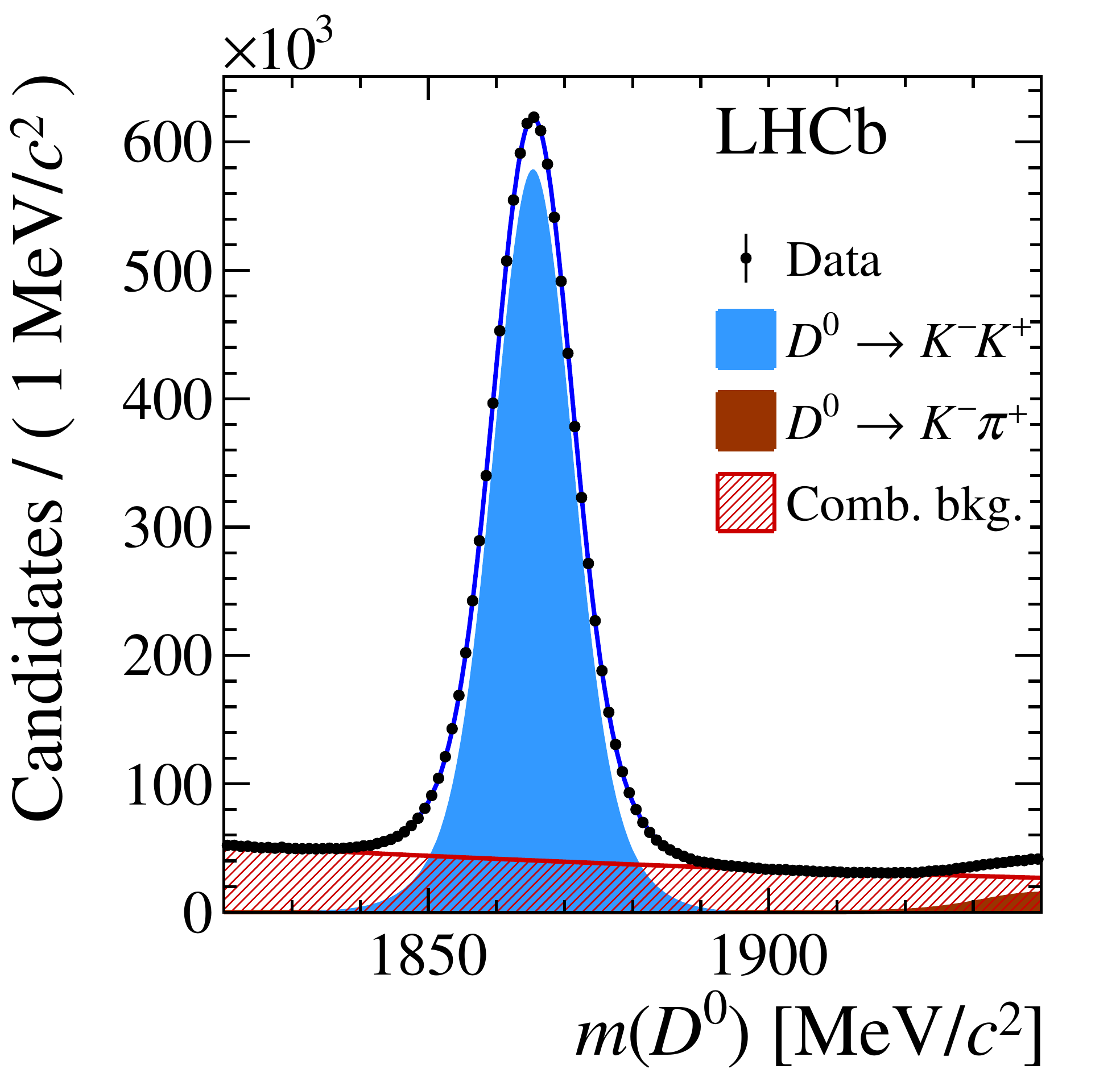}
\includegraphics[width=0.45\textwidth]{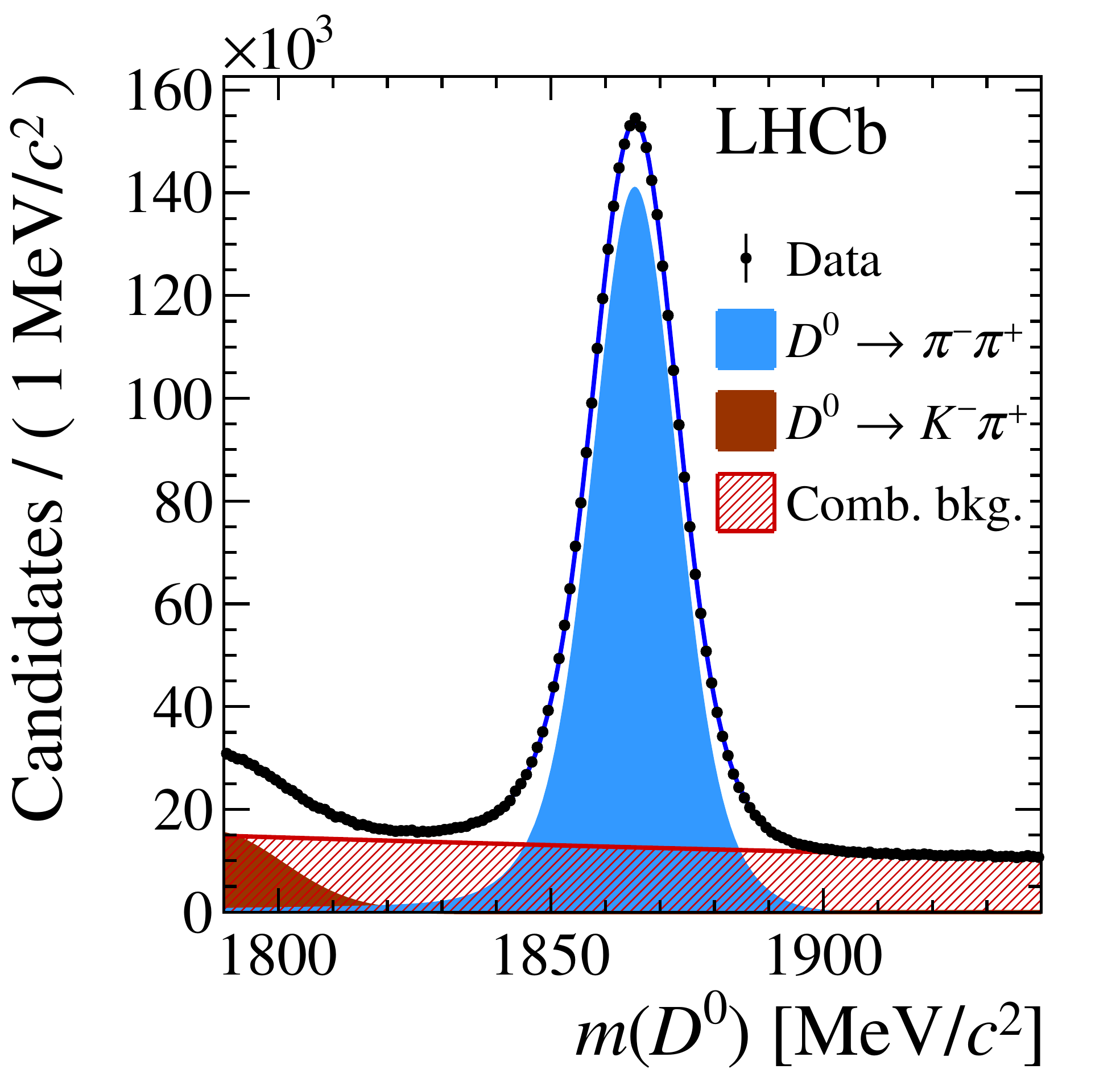}
\caption{Mass distributions of selected (top)~$\pi^\pm$-tagged and (bottom)~$\mu^\pm$-tagged candidates for (left)~$\Km\Kp$ and (right)~$\pim\pip$ final states of the \Dz-meson decays, with fit projections overlaid.}
\label{fig:fits}
\end{figure}

\section{Measurement of the Asymmetries}

For each decay mode, simultaneous least-square fits to the binned mass distributions of \Dstarp and \Dstarm candidates for the $\pi$-tagged sample, or \Dz and \Dzb candidates for the $\mu$-tagged sample, are performed to obtain the raw asymmetries of signal and background components, which are free parameters of the fits.

In the analysis of the $\pi$-tagged sample the fits are performed to the $m(\Dz\pip)$ and $m(\Dzb\pim)$ distributions, that are defined using the known value of the \Dz mass~\cite{bib:cdfpaper}.
The signal mass model consists of the sum of three Gaussian functions and a Johnson $S_{{U}}$ function~\cite{Johnson:1949zj}, whereas the combinatorial background is described by an empirical function of the form $[m(\Dz\pip) - m(\Dz) - m(\pip)]^\alpha e^{\beta\, m(\Dz\pip)}$.
All the parameters of the models are free to be adjusted by the fit and are shared among positive and negative tags, except for the mean values of the Gaussian functions, which are different to take into account small shifts in the raw mass measurements between opposite tags.

In the analysis of the $\mu$-tagged sample, the fits are performed to the $m(\Dz)$ distributions.
The signal is described by the sum of two Gaussian functions convolved with a truncated power-law function accounting for final-state photon radiation effects, while the combinatorial background is modeled by an exponential function.
A small contribution from $\decay{\Dz}{\Km\pip}$ decays with a misidentified kaon or pion is visible and is modeled as the tail of a Gaussian function.
The fit parameters are shared among positive and negative tags, except for the mean values of the Gaussian functions.

Fits are performed to subsamples of data split according to magnet polarities and years of data taking.
The final results are obtained by averaging the partial \DACP values corresponding to each subsample, which are found to be in good agreement.
Performing single fits the overall $\pi$-tagged and $\mu$-tagged samples gives small differences of the order of a few $10^{-5}$.
Figure~\ref{fig:fits} displays the $m(\Dz\pip)$ and $m(\Dz)$ distributions corresponding to the entire samples.
The $\pi$-tagged~($\mu$-tagged) signal yields are approximately $44$~($9$) million  $\dkk$ decays and $14$~($3$) million $\D^0 \rightarrow \pi^-\pi^+$ decays.

\section{Systematic Uncertainties}

Several sources of systematic uncertainties affecting the measurement are considered and studied independently for the $\pi$-tagged and $\mu$-tagged samples.
In the case of $\pi$-tagged decays, the dominant systematic uncertainty is related to the knowledge of the signal and background mass models.
It is evaluated by generating pseudoexperiments according to the baseline fit model, then fitting both baseline and alternative models to those data and considering the difference between the resulting values of \DACP.
A value of $0.6\times 10^{-4}$, corresponding to the largest observed variation, is assigned as a systematic uncertainty. 
A similar study with pseudoexperiments is also performed with the $\mu$-tagged sample and a value of $2\times 10^{-4}$ is found.

In the case of $\mu$-tagged decays, the main systematic uncertainty is due to the possibility that the \Dz flavor is not tagged correctly by the muon charge because of misreconstruction.
The probability of wrongly assigning the \Dz flavor~(mistag) is measured on a large sample of $\mu$-tagged \dkpi decays by comparing the charges of kaon and muon candidates.
Mistag rates are found to be at the percent level and compatible for positively and negatively tagged decays, and the corresponding systematic uncertainty is estimated to be $4\times 10^{-4}$.

Systematic uncertainties of $0.2\times 10^{-4}$ and $1\times 10^{-4}$ accounting for the knowledge of the weights used in the kinematic weighting procedure are assessed for $\pi$-tagged and $\mu$-tagged decays, respectively.
A fraction of \Dz mesons from $B$ decays~(secondary decays) is still present in the final $\pi$-tagged sample even after the requirement that the \Dz trajectory points back to the PV.
Possible different levels of contamination from secondary decays in \dkk and \dpipi samples may bias the value of \DACP because of an incomplete cancellation of the production asymmetries of $b$ hadrons.
The fractions of secondary decays are estimated by performing a fit to the distribution of the \Dz-candidate impact parameter in the plane transverse to the beam direction, and the corresponding systematic uncertainty is estimated to be $0.3\times 10^{-4}$.
A systematic uncertainty associated to the presence of background components peaking in $m(\Dz\pi)$ and not in $m(\Dz)$ is determined by fits to the $m(\Dz)$ distributions after the removal of the signal window requirement, where these components are modeled using fast simulation.
The main sources are the $\decay{\Dz}{\Km\pip\piz}$ decay for the $\Kp\Km$ mode, and the $\Dz\to\pim\mup\neum$ and $\Dz\to\pim\ep\neue$ decays for the $\pip\pim$ mode.
Yields and raw asymmetries of the peaking-background components measured from the fits are then used as inputs to pseudoexperiments performed to evaluate the corresponding effects on the determination of \DACP, resulting in a systematic uncertainty of $0.5\times 10^{-4}$.

In the case of $\mu$-tagged decays, the fractions of reconstructed \Bb decays can be slightly different between the $K^-K^+$ and $\pi^-\pi^+$ decay modes, which could lead to a small bias in \DACP.
Using the LHCb measurements of the $b$-hadron production asymmetries~\cite{LHCb-PAPER-2016-062}, the associated systematic uncertainty on \DACP is estimated to be $1\times 10^{-4}$.
The combination of a difference in the \B reconstruction efficiency as a function of the decay time between the \dkk and \dpipi modes and the presence of neutral $B$-meson oscillations may also cause an imperfect cancellation of \APB in \DACP, and the related systematic uncertainty is estimated to be $2\times 10^{-4}$.

The total systematic uncertainties on \DACP are given by the sum in quadrature of all individual contributions, and are equal to $0.9\times 10^{-4}$ and $5\times 10^{-4}$ for the $\pi$-tagged and $\mu$-tagged samples, respectively.
A summary of all systematic uncertainties is reported in Table~\ref{tab:syst}.

\begin{table}[t]
\centering
\caption{Systematic uncertainties on \DACP for $\pi$- and $\mu$-tagged decays (in $10^{-4}$). The total uncertainties are obtained as the sums in quadrature of the individual contributions.}
\label{tab:syst}
        \begin{tabular}{| lcc |}
            \hline    
                Source & $\pi$-tagged & $\mu$-tagged \\
                \hline
                Fit model                        & 0.6         & 2  \\
                Mistag                           & --          & 4\\
                Weighting                        & 0.2         & 1 \\
                Secondary decays                 & 0.3         & -- \\
                Peaking background               & 0.5         & --\\
                \B fractions                     & --          & 1\\
                \B reco. efficiency              & --          & 2 \\
                \hline
                Total                            & 0.9         & 5\\
                \hline
        \end{tabular}
\end{table}

Numerous additional robustness checks are carried out.
The measured value of $\Delta A_{\CP}$ is studied as a function of several geometrical and kinematic variables.
Furthermore, the total sample is split into subsamples taken in different run periods within the years of data taking, also distinguishing different magnet polarities.
No evidence for unexpected dependences of \DACP is found in any of these tests.
A check using more stringent PID requirements is performed, and all variations of \DACP are found to be compatible within statistical uncertainties.
An additional check concerns the measurement of $\Delta A_{\rm bkg}$, which is the difference of the background raw asymmetries in $\Km\Kp$ and $\pim\pip$ final states.
The prompt background is mainly composed of genuine \Dz candidates paired with unrelated pions originating from the PV, so $\Delta A_{\rm bkg}$ is expected to be compatible with zero.
A value of $\Delta A_{\rm bkg} = (-2 \pm 4) \times 10^{-4}$ is obtained.

\section{Results}

The measured differences of time-integrated \CP asymmetries of \decay{\Dz}{\Km\Kp} and $\Dz\to\pim\pip$ decays are~\cite{LHCb-PAPER-2019-006}
\begin{align*}
	\DACP^{\pi{\mbox{-}}\rm{tagged}} &= \left[-18.2 \pm 3.2\,\rm{(stat.)} \pm 0.9\,\rm{(syst.)} \right]\times 10^{-4},\\ 
	\DACP^{\mu{\mbox{-}}\rm{tagged}} &= \left[-9 \pm 8\,\rm{(stat.)}  \pm 5\,\rm{(syst.)} \right] \times 10^{-4},
\end{align*}
both in good agreement with world averages~\cite{HFLAV16} and previous LHCb results~\cite{LHCB-PAPER-2015-055,LHCB-PAPER-2014-013}. 

The full combination with previous LHCb measurements~\cite{LHCB-PAPER-2015-055,LHCB-PAPER-2014-013} gives the following value of \DACP
\begin{equation}
\DACP = \left(-15.4 \pm 2.9 \right)\times 10^{-4}, \nonumber
\end{equation}
where the uncertainty includes statistical and systematic contributions.
The significance of the deviation from zero corresponds to 5.3 standard deviations.
This is the first observation of \CP violation in the decay of charm hadrons.

As shown in Eq.~\ref{eq:dacpdef}, the interpretation of \DACP in terms of direct \CP violation and $A_\Gamma$ requires knowledge of the difference of reconstructed mean decay times for \dkk and \dpipi decays normalized to the \Dz lifetime.
The values corresponding to the present measurements, using the world average of the \Dz lifetime~\cite{PDG2018}, are $\Delta\!\mean{t}^{\pi{\mbox{-}}\rm{tagged}}/\tau(\Dz)  =  0.135 \pm 0.002$ and $\Delta\!\mean{t}^{\mu{\mbox{-}}\rm{tagged}}/\tau(\Dz)  =  -0.003 \pm 0.001$, whereas that corresponding to the full combination is $\Delta\mean{t}/\tau(\Dz)  =  0.115 \pm 0.002$.
The uncertainties include statistical and systematic contributions.
By using the LHCb average~\cite{LHCb-PAPER-2014-069,LHCb-PAPER-2016-063} \mbox{$\agamma=(-2.8 \pm 2.8)\times 10^{-4}$}, from Eq.~\ref{eq:dacpdef} it is possible to derive
\begin{equation*}
	\Delta a_{\CP}^{\rm dir} = \left(-15.7 \pm 2.9 \right)\times 10^{-4}.
\end{equation*}
As expected, \DACP is primarily sensitive to direct \CP violation.

In summary, this document reports the first observation of a nonzero \CP asymmetry in charm decays, using large samples of \dkk and \dpipi decays collected with the LHCb detector.
The result is consistent with, although in magnitude at the upper end of, SM expectations.
In the next future, further measurements with charmed particles, along with possible theoretical improvements, will help clarify the present physics picture, to establish whether this result is consistent with the SM or indicates the presence of new physics processes in the up-quark sector.

\section*{References}

\setboolean{inbibliography}{true}
\bibliography{betti_federico}

\ifx\mcitethebibliography\mciteundefinedmacro
\PackageError{LHCb.bst}{mciteplus.sty has not been loaded}
{This bibstyle requires the use of the mciteplus package.}\fi
\providecommand{\href}[2]{#2}
\begin{mcitethebibliography}{10}
\mciteSetBstSublistMode{n}
\mciteSetBstMaxWidthForm{subitem}{\alph{mcitesubitemcount})}
\mciteSetBstSublistLabelBeginEnd{\mcitemaxwidthsubitemform\space}
{\relax}{\relax}

\bibitem{Sakharov:1967dj}
A.~D. Sakharov, \ifthenelse{\boolean{articletitles}}{\emph{{Violation of CP
  invariance, C asymmetry, and baryon asymmetry of the universe}},
  }{}\href{https://doi.org/10.1070/PU1991v034n05ABEH002497}{Pisma Zh.\ Eksp.\
  Teor.\ Fiz.\  \textbf{5} (1967) 32}\relax
\mciteBstWouldAddEndPuncttrue
\mciteSetBstMidEndSepPunct{\mcitedefaultmidpunct}
{\mcitedefaultendpunct}{\mcitedefaultseppunct}\relax
\EndOfBibitem
\bibitem{Cabibbo:1963yz}
N.~Cabibbo, \ifthenelse{\boolean{articletitles}}{\emph{{Unitary symmetry and
  leptonic decays}},
  }{}\href{https://doi.org/10.1103/PhysRevLett.10.531}{Phys.\ Rev.\ Lett.\
  \textbf{10} (1963) 531}\relax
\mciteBstWouldAddEndPuncttrue
\mciteSetBstMidEndSepPunct{\mcitedefaultmidpunct}
{\mcitedefaultendpunct}{\mcitedefaultseppunct}\relax
\EndOfBibitem
\bibitem{Kobayashi:1973fv}
M.~Kobayashi and T.~Maskawa,
  \ifthenelse{\boolean{articletitles}}{\emph{{\CP-violation in the
  renormalizable theory of weak interaction}},
  }{}\href{https://doi.org/10.1143/PTP.49.652}{Prog.\ Theor.\ Phys.\
  \textbf{49} (1973) 652}\relax
\mciteBstWouldAddEndPuncttrue
\mciteSetBstMidEndSepPunct{\mcitedefaultmidpunct}
{\mcitedefaultendpunct}{\mcitedefaultseppunct}\relax
\EndOfBibitem
\bibitem{Christenson:1964fg}
J.~H. Christenson, J.~W. Cronin, V.~L. Fitch, and R.~Turlay,
  \ifthenelse{\boolean{articletitles}}{\emph{{Evidence for the $2\pi$ decay of
  the $K_2^0$ meson}},
  }{}\href{https://doi.org/10.1103/PhysRevLett.13.138}{Phys.\ Rev.\ Lett.\
  \textbf{13} (1964) 138}\relax
\mciteBstWouldAddEndPuncttrue
\mciteSetBstMidEndSepPunct{\mcitedefaultmidpunct}
{\mcitedefaultendpunct}{\mcitedefaultseppunct}\relax
\EndOfBibitem
\bibitem{AlaviHarati:1999xp}
KTeV collaboration, A.~Alavi-Harati {\em et~al.},
  \ifthenelse{\boolean{articletitles}}{\emph{{Observation of direct CP
  violation in $K_{\rm S,L} \to \pi \pi$ decays}},
  }{}\href{https://doi.org/10.1103/PhysRevLett.83.22}{Phys.\ Rev.\ Lett.\
  \textbf{83} (1999) 22},
  \href{http://arxiv.org/abs/hep-ex/9905060}{{\normalfont\ttfamily
  arXiv:hep-ex/9905060}}\relax
\mciteBstWouldAddEndPuncttrue
\mciteSetBstMidEndSepPunct{\mcitedefaultmidpunct}
{\mcitedefaultendpunct}{\mcitedefaultseppunct}\relax
\EndOfBibitem
\bibitem{Lai:2001ki}
NA48 collaboration, A.~Lai {\em et~al.},
  \ifthenelse{\boolean{articletitles}}{\emph{{A precise measurement of the
  direct CP violation parameter $Re(\eps^\prime / \eps)$}},
  }{}\href{https://doi.org/10.1007/s100520100822}{Eur.\ Phys.\ J.\
  \textbf{C22} (2001) 231},
  \href{http://arxiv.org/abs/hep-ex/0110019}{{\normalfont\ttfamily
  arXiv:hep-ex/0110019}}\relax
\mciteBstWouldAddEndPuncttrue
\mciteSetBstMidEndSepPunct{\mcitedefaultmidpunct}
{\mcitedefaultendpunct}{\mcitedefaultseppunct}\relax
\EndOfBibitem
\bibitem{Aubert:2001nu}
BaBar collaboration, B.~Aubert {\em et~al.},
  \ifthenelse{\boolean{articletitles}}{\emph{{Observation of \CP violation in
  the $B^0$ meson system}},
  }{}\href{https://doi.org/10.1103/PhysRevLett.87.091801}{Phys.\ Rev.\ Lett.\
  \textbf{87} (2001) 091801},
  \href{http://arxiv.org/abs/hep-ex/0107013}{{\normalfont\ttfamily
  arXiv:hep-ex/0107013}}\relax
\mciteBstWouldAddEndPuncttrue
\mciteSetBstMidEndSepPunct{\mcitedefaultmidpunct}
{\mcitedefaultendpunct}{\mcitedefaultseppunct}\relax
\EndOfBibitem
\bibitem{Abe:2001xe}
Belle collaboration, K.~Abe {\em et~al.},
  \ifthenelse{\boolean{articletitles}}{\emph{{Observation of large \CP
  violation in the neutral $B$ meson system}},
  }{}\href{https://doi.org/10.1103/PhysRevLett.87.091802}{Phys.\ Rev.\ Lett.\
  \textbf{87} (2001) 091802},
  \href{http://arxiv.org/abs/hep-ex/0107061}{{\normalfont\ttfamily
  arXiv:hep-ex/0107061}}\relax
\mciteBstWouldAddEndPuncttrue
\mciteSetBstMidEndSepPunct{\mcitedefaultmidpunct}
{\mcitedefaultendpunct}{\mcitedefaultseppunct}\relax
\EndOfBibitem
\bibitem{Aubert:2004qm}
BaBar collaboration, B.~Aubert {\em et~al.},
  \ifthenelse{\boolean{articletitles}}{\emph{{Direct \CP violating asymmetry in
  \mbox{$B^0 \to K^+\pi^-$} decays}},
  }{}\href{https://doi.org/10.1103/PhysRevLett.93.131801}{Phys.\ Rev.\ Lett.\
  \textbf{93} (2004) 131801},
  \href{http://arxiv.org/abs/hep-ex/0407057}{{\normalfont\ttfamily
  arXiv:hep-ex/0407057}}\relax
\mciteBstWouldAddEndPuncttrue
\mciteSetBstMidEndSepPunct{\mcitedefaultmidpunct}
{\mcitedefaultendpunct}{\mcitedefaultseppunct}\relax
\EndOfBibitem
\bibitem{Chao:2004mn}
Belle collaboration, Y.~Chao {\em et~al.},
  \ifthenelse{\boolean{articletitles}}{\emph{{Evidence for direct CP violation
  in $B^0\to \Kp \pim$ decays}},
  }{}\href{https://doi.org/10.1103/PhysRevLett.93.191802}{Phys.\ Rev.\ Lett.\
  \textbf{93} (2004) 191802},
  \href{http://arxiv.org/abs/hep-ex/0408100}{{\normalfont\ttfamily
  arXiv:hep-ex/0408100}}\relax
\mciteBstWouldAddEndPuncttrue
\mciteSetBstMidEndSepPunct{\mcitedefaultmidpunct}
{\mcitedefaultendpunct}{\mcitedefaultseppunct}\relax
\EndOfBibitem
\bibitem{LHCb-PAPER-2013-018}
LHCb collaboration, R.~Aaij {\em et~al.},
  \ifthenelse{\boolean{articletitles}}{\emph{{First observation of \CP
  violation in the decays of $\Bs$ mesons}},
  }{}\href{https://doi.org/10.1103/PhysRevLett.110.221601}{Phys.\ Rev.\ Lett.\
  \textbf{110} (2013) 221601},
  \href{http://arxiv.org/abs/1304.6173}{{\normalfont\ttfamily
  arXiv:1304.6173}}\relax
\mciteBstWouldAddEndPuncttrue
\mciteSetBstMidEndSepPunct{\mcitedefaultmidpunct}
{\mcitedefaultendpunct}{\mcitedefaultseppunct}\relax
\EndOfBibitem
\bibitem{LHCb-PAPER-2012-001}
LHCb collaboration, R.~Aaij {\em et~al.},
  \ifthenelse{\boolean{articletitles}}{\emph{{Observation of \CP violation in
  $\Bpm\to \D\Kpm$ decays}},
  }{}\href{https://doi.org/10.1016/j.physletb.2012.04.060}{Phys.\ Lett.\
  \textbf{B712} (2012) 203}, Erratum
  \href{https://doi.org/10.1016/j.physletb.2012.05.060}{ibid.\   \textbf{B713}
  (2012) 351}, \href{http://arxiv.org/abs/1203.3662}{{\normalfont\ttfamily
  arXiv:1203.3662}}\relax
\mciteBstWouldAddEndPuncttrue
\mciteSetBstMidEndSepPunct{\mcitedefaultmidpunct}
{\mcitedefaultendpunct}{\mcitedefaultseppunct}\relax
\EndOfBibitem
\bibitem{Cohen:1993nk}
A.~G. Cohen, D.~B. Kaplan, and A.~E. Nelson,
  \ifthenelse{\boolean{articletitles}}{\emph{{Progress in electroweak
  baryogenesis}},
  }{}\href{https://doi.org/10.1146/annurev.ns.43.120193.000331}{Ann.\ Rev.\
  Nucl.\ Part.\ Sci.\  \textbf{43} (1993) 27},
  \href{http://arxiv.org/abs/hep-ph/9302210}{{\normalfont\ttfamily
  arXiv:hep-ph/9302210}}\relax
\mciteBstWouldAddEndPuncttrue
\mciteSetBstMidEndSepPunct{\mcitedefaultmidpunct}
{\mcitedefaultendpunct}{\mcitedefaultseppunct}\relax
\EndOfBibitem
\bibitem{Riotto:1999yt}
A.~Riotto and M.~Trodden, \ifthenelse{\boolean{articletitles}}{\emph{{Recent
  progress in baryogenesis}},
  }{}\href{https://doi.org/10.1146/annurev.nucl.49.1.35}{Ann.\ Rev.\ Nucl.\
  Part.\ Sci.\  \textbf{49} (1999) 35},
  \href{http://arxiv.org/abs/hep-ph/9901362}{{\normalfont\ttfamily
  arXiv:hep-ph/9901362}}\relax
\mciteBstWouldAddEndPuncttrue
\mciteSetBstMidEndSepPunct{\mcitedefaultmidpunct}
{\mcitedefaultendpunct}{\mcitedefaultseppunct}\relax
\EndOfBibitem
\bibitem{Hou:2008xd}
W.-S. Hou, \ifthenelse{\boolean{articletitles}}{\emph{{Source of CP violation
  for the baryon asymmetry of the universe}},
  }{}\href{https://www.ps-taiwan.org/cjp/issues.php?vol=47&num=2}{Chin.\ J.\
  Phys.\  \textbf{47} (2009) 134},
  \href{http://arxiv.org/abs/0803.1234}{{\normalfont\ttfamily
  arXiv:0803.1234}}\relax
\mciteBstWouldAddEndPuncttrue
\mciteSetBstMidEndSepPunct{\mcitedefaultmidpunct}
{\mcitedefaultendpunct}{\mcitedefaultseppunct}\relax
\EndOfBibitem
\bibitem{Golden:1989qx}
M.~Golden and B.~Grinstein,
  \ifthenelse{\boolean{articletitles}}{\emph{{Enhanced CP violations in
  hadronic charm decays}},
  }{}\href{https://doi.org/10.1016/0370-2693(89)90353-5}{Phys.\ Lett.\
  \textbf{B222} (1989) 501}\relax
\mciteBstWouldAddEndPuncttrue
\mciteSetBstMidEndSepPunct{\mcitedefaultmidpunct}
{\mcitedefaultendpunct}{\mcitedefaultseppunct}\relax
\EndOfBibitem
\bibitem{Buccella:1994nf}
F.~Buccella {\em et~al.},
  \ifthenelse{\boolean{articletitles}}{\emph{{Nonleptonic weak decays of
  charmed mesons}}, }{}\href{https://doi.org/10.1103/PhysRevD.51.3478}{Phys.\
  Rev.\  \textbf{D51} (1995) 3478},
  \href{http://arxiv.org/abs/hep-ph/9411286}{{\normalfont\ttfamily
  arXiv:hep-ph/9411286}}\relax
\mciteBstWouldAddEndPuncttrue
\mciteSetBstMidEndSepPunct{\mcitedefaultmidpunct}
{\mcitedefaultendpunct}{\mcitedefaultseppunct}\relax
\EndOfBibitem
\bibitem{Bianco:2003vb}
S.~Bianco, F.~L. Fabbri, D.~Benson, and I.~Bigi,
  \ifthenelse{\boolean{articletitles}}{\emph{{A Cicerone for the physics of
  charm}}, }{}\href{https://doi.org/10.1393/ncr/i2003-10003-1}{Riv.\ Nuovo
  Cim.\  \textbf{26N7} (2003) 1},
  \href{http://arxiv.org/abs/hep-ex/0309021}{{\normalfont\ttfamily
  arXiv:hep-ex/0309021}}\relax
\mciteBstWouldAddEndPuncttrue
\mciteSetBstMidEndSepPunct{\mcitedefaultmidpunct}
{\mcitedefaultendpunct}{\mcitedefaultseppunct}\relax
\EndOfBibitem
\bibitem{Grossman:2006jg}
Y.~Grossman, A.~L. Kagan, and Y.~Nir,
  \ifthenelse{\boolean{articletitles}}{\emph{{New physics and CP violation in
  singly Cabibbo suppressed D decays}},
  }{}\href{https://doi.org/10.1103/PhysRevD.75.036008}{Phys.\ Rev.\
  \textbf{D75} (2007) 036008},
  \href{http://arxiv.org/abs/hep-ph/0609178}{{\normalfont\ttfamily
  arXiv:hep-ph/0609178}}\relax
\mciteBstWouldAddEndPuncttrue
\mciteSetBstMidEndSepPunct{\mcitedefaultmidpunct}
{\mcitedefaultendpunct}{\mcitedefaultseppunct}\relax
\EndOfBibitem
\bibitem{Artuso:2008vf}
M.~Artuso, B.~Meadows, and A.~A. Petrov,
  \ifthenelse{\boolean{articletitles}}{\emph{{Charm meson decays}},
  }{}\href{https://doi.org/10.1146/annurev.nucl.58.110707.171131}{Ann.\ Rev.\
  Nucl.\ Part.\ Sci.\  \textbf{58} (2008) 249},
  \href{http://arxiv.org/abs/0802.2934}{{\normalfont\ttfamily
  arXiv:0802.2934}}\relax
\mciteBstWouldAddEndPuncttrue
\mciteSetBstMidEndSepPunct{\mcitedefaultmidpunct}
{\mcitedefaultendpunct}{\mcitedefaultseppunct}\relax
\EndOfBibitem
\bibitem{Brod:2011re}
J.~Brod, A.~L. Kagan, and J.~Zupan,
  \ifthenelse{\boolean{articletitles}}{\emph{{Size of direct CP violation in
  singly Cabibbo-suppressed D decays}},
  }{}\href{https://doi.org/10.1103/PhysRevD.86.014023}{Phys.\ Rev.\
  \textbf{D86} (2012) 014023},
  \href{http://arxiv.org/abs/1111.5000}{{\normalfont\ttfamily
  arXiv:1111.5000}}\relax
\mciteBstWouldAddEndPuncttrue
\mciteSetBstMidEndSepPunct{\mcitedefaultmidpunct}
{\mcitedefaultendpunct}{\mcitedefaultseppunct}\relax
\EndOfBibitem
\bibitem{Cheng:2012wr}
H.-Y. Cheng and C.-W. Chiang,
  \ifthenelse{\boolean{articletitles}}{\emph{{Direct \CP violation in two-body
  hadronic charmed meson decays}},
  }{}\href{https://doi.org/10.1103/PhysRevD.85.034036}{Phys.\ Rev.\
  \textbf{D85} (2012) 034036}, Erratum
  \href{https://doi.org/10.1103/PhysRevD.85.079903}{ibid.\   \textbf{D85}
  (2012) 079903}, \href{http://arxiv.org/abs/1201.0785}{{\normalfont\ttfamily
  arXiv:1201.0785}}\relax
\mciteBstWouldAddEndPuncttrue
\mciteSetBstMidEndSepPunct{\mcitedefaultmidpunct}
{\mcitedefaultendpunct}{\mcitedefaultseppunct}\relax
\EndOfBibitem
\bibitem{Cheng:2012xb}
H.-Y. Cheng and C.-W. Chiang, \ifthenelse{\boolean{articletitles}}{\emph{{SU(3)
  symmetry breaking and \CP violation in $D \to PP$ decays}},
  }{}\href{https://doi.org/10.1103/PhysRevD.86.014014}{Phys.\ Rev.\
  \textbf{D86} (2012) 014014},
  \href{http://arxiv.org/abs/1205.0580}{{\normalfont\ttfamily
  arXiv:1205.0580}}\relax
\mciteBstWouldAddEndPuncttrue
\mciteSetBstMidEndSepPunct{\mcitedefaultmidpunct}
{\mcitedefaultendpunct}{\mcitedefaultseppunct}\relax
\EndOfBibitem
\bibitem{Li:2012cfa}
H.-n. Li, C.-D. Lu, and F.-S. Yu,
  \ifthenelse{\boolean{articletitles}}{\emph{{Branching ratios and direct CP
  asymmetries in $D\to PP$ decays}},
  }{}\href{https://doi.org/10.1103/PhysRevD.86.036012}{Phys.\ Rev.\
  \textbf{D86} (2012) 036012},
  \href{http://arxiv.org/abs/1203.3120}{{\normalfont\ttfamily
  arXiv:1203.3120}}\relax
\mciteBstWouldAddEndPuncttrue
\mciteSetBstMidEndSepPunct{\mcitedefaultmidpunct}
{\mcitedefaultendpunct}{\mcitedefaultseppunct}\relax
\EndOfBibitem
\bibitem{Franco:2012ck}
E.~Franco, S.~Mishima, and L.~Silvestrini,
  \ifthenelse{\boolean{articletitles}}{\emph{{The Standard Model confronts \CP
  violation in $D^0 \to \pi^+\pi^-$ and $D^0 \to K^+K^-$}},
  }{}\href{https://doi.org/10.1007/JHEP05(2012)140}{JHEP \textbf{05} (2012)
  140}, \href{http://arxiv.org/abs/1203.3131}{{\normalfont\ttfamily
  arXiv:1203.3131}}\relax
\mciteBstWouldAddEndPuncttrue
\mciteSetBstMidEndSepPunct{\mcitedefaultmidpunct}
{\mcitedefaultendpunct}{\mcitedefaultseppunct}\relax
\EndOfBibitem
\bibitem{Pirtskhalava:2011va}
D.~Pirtskhalava and P.~Uttayarat,
  \ifthenelse{\boolean{articletitles}}{\emph{{CP Violation and flavor SU(3)
  breaking in D-meson decays}},
  }{}\href{https://doi.org/10.1016/j.physletb.2012.04.039}{Phys.\ Lett.\
  \textbf{B712} (2012) 81},
  \href{http://arxiv.org/abs/1112.5451}{{\normalfont\ttfamily
  arXiv:1112.5451}}\relax
\mciteBstWouldAddEndPuncttrue
\mciteSetBstMidEndSepPunct{\mcitedefaultmidpunct}
{\mcitedefaultendpunct}{\mcitedefaultseppunct}\relax
\EndOfBibitem
\bibitem{Feldmann:2012js}
T.~Feldmann, S.~Nandi, and A.~Soni,
  \ifthenelse{\boolean{articletitles}}{\emph{{Repercussions of flavour symmetry
  breaking on CP violation in D-meson decays}},
  }{}\href{https://doi.org/10.1007/JHEP06(2012)007}{JHEP \textbf{06} (2012)
  007}, \href{http://arxiv.org/abs/1202.3795}{{\normalfont\ttfamily
  arXiv:1202.3795}}\relax
\mciteBstWouldAddEndPuncttrue
\mciteSetBstMidEndSepPunct{\mcitedefaultmidpunct}
{\mcitedefaultendpunct}{\mcitedefaultseppunct}\relax
\EndOfBibitem
\bibitem{Brod:2012ud}
J.~Brod, Y.~Grossman, A.~L. Kagan, and J.~Zupan,
  \ifthenelse{\boolean{articletitles}}{\emph{{A consistent picture for large
  penguins in $D \to \pip\pim,\,\Kp\Km$}},
  }{}\href{https://doi.org/10.1007/JHEP10(2012)161}{JHEP \textbf{10} (2012)
  161}, \href{http://arxiv.org/abs/1203.6659}{{\normalfont\ttfamily
  arXiv:1203.6659}}\relax
\mciteBstWouldAddEndPuncttrue
\mciteSetBstMidEndSepPunct{\mcitedefaultmidpunct}
{\mcitedefaultendpunct}{\mcitedefaultseppunct}\relax
\EndOfBibitem
\bibitem{Hiller:2012xm}
G.~Hiller, M.~Jung, and S.~Schacht,
  \ifthenelse{\boolean{articletitles}}{\emph{{SU(3)-flavor anatomy of
  nonleptonic charm decays}},
  }{}\href{https://doi.org/10.1103/PhysRevD.87.014024}{Phys.\ Rev.\
  \textbf{D87} (2013) 014024},
  \href{http://arxiv.org/abs/1211.3734}{{\normalfont\ttfamily
  arXiv:1211.3734}}\relax
\mciteBstWouldAddEndPuncttrue
\mciteSetBstMidEndSepPunct{\mcitedefaultmidpunct}
{\mcitedefaultendpunct}{\mcitedefaultseppunct}\relax
\EndOfBibitem
\bibitem{Grossman:2012ry}
Y.~Grossman and D.~J. Robinson,
  \ifthenelse{\boolean{articletitles}}{\emph{{SU(3) sum rules for charm
  decay}}, }{}\href{https://doi.org/10.1007/JHEP04(2013)067}{JHEP \textbf{04}
  (2013) 067}, \href{http://arxiv.org/abs/1211.3361}{{\normalfont\ttfamily
  arXiv:1211.3361}}\relax
\mciteBstWouldAddEndPuncttrue
\mciteSetBstMidEndSepPunct{\mcitedefaultmidpunct}
{\mcitedefaultendpunct}{\mcitedefaultseppunct}\relax
\EndOfBibitem
\bibitem{Bhattacharya:2012ah}
B.~Bhattacharya, M.~Gronau, and J.~L. Rosner,
  \ifthenelse{\boolean{articletitles}}{\emph{{CP asymmetries in
  singly-Cabibbo-suppressed $D$ decays to two pseudoscalar mesons}},
  }{}\href{https://doi.org/10.1103/PhysRevD.85.079901}{Phys.\ Rev.\
  \textbf{D85} (2012) 054014},
  \href{http://arxiv.org/abs/1201.2351}{{\normalfont\ttfamily
  arXiv:1201.2351}}\relax
\mciteBstWouldAddEndPuncttrue
\mciteSetBstMidEndSepPunct{\mcitedefaultmidpunct}
{\mcitedefaultendpunct}{\mcitedefaultseppunct}\relax
\EndOfBibitem
\bibitem{Muller:2015rna}
S.~M{\"u}ller, U.~Nierste, and S.~Schacht,
  \ifthenelse{\boolean{articletitles}}{\emph{{Sum rules of charm \CP
  asymmetries beyond the SU(3)$_F$ limit}},
  }{}\href{https://doi.org/10.1103/PhysRevLett.115.251802}{Phys.\ Rev.\ Lett.\
  \textbf{115} (2015) 251802},
  \href{http://arxiv.org/abs/1506.04121}{{\normalfont\ttfamily
  arXiv:1506.04121}}\relax
\mciteBstWouldAddEndPuncttrue
\mciteSetBstMidEndSepPunct{\mcitedefaultmidpunct}
{\mcitedefaultendpunct}{\mcitedefaultseppunct}\relax
\EndOfBibitem
\bibitem{Khodjamirian:2017zdu}
A.~Khodjamirian and A.~A. Petrov,
  \ifthenelse{\boolean{articletitles}}{\emph{{Direct CP asymmetry in $D\to
  \pi^-\pi^+$ and {\mbox{$D\to K^-K^+$}} in QCD-based approach}},
  }{}\href{https://doi.org/10.1016/j.physletb.2017.09.070}{Phys.\ Lett.\
  \textbf{B774} (2017) 235},
  \href{http://arxiv.org/abs/1706.07780}{{\normalfont\ttfamily
  arXiv:1706.07780}}\relax
\mciteBstWouldAddEndPuncttrue
\mciteSetBstMidEndSepPunct{\mcitedefaultmidpunct}
{\mcitedefaultendpunct}{\mcitedefaultseppunct}\relax
\EndOfBibitem
\bibitem{Buccella:2019kpn}
F.~Buccella, A.~Paul, and P.~Santorelli,
  \ifthenelse{\boolean{articletitles}}{\emph{{On $SU(3)_{F}$ breaking through
  final state interactions and CP asymmetries in $D\to P P$ decays}},
  }{}\href{http://arxiv.org/abs/1902.05564}{{\normalfont\ttfamily
  arXiv:1902.05564}}\relax
\mciteBstWouldAddEndPuncttrue
\mciteSetBstMidEndSepPunct{\mcitedefaultmidpunct}
{\mcitedefaultendpunct}{\mcitedefaultseppunct}\relax
\EndOfBibitem
\bibitem{bib:babarpaper2008}
BaBar collaboration, B.~Aubert {\em et~al.},
  \ifthenelse{\boolean{articletitles}}{\emph{{Search for CP violation in the
  decays {\mbox{$D^0 \to K^{-}K^{+}$}} and $D^0 \to \pi^{-} \pi^{+}$}},
  }{}\href{https://doi.org/10.1103/PhysRevLett.100.061803}{Phys.\ Rev.\ Lett.\
  \textbf{100} (2008) 061803},
  \href{http://arxiv.org/abs/0709.2715}{{\normalfont\ttfamily
  arXiv:0709.2715}}\relax
\mciteBstWouldAddEndPuncttrue
\mciteSetBstMidEndSepPunct{\mcitedefaultmidpunct}
{\mcitedefaultendpunct}{\mcitedefaultseppunct}\relax
\EndOfBibitem
\bibitem{bib:bellepaper2008}
Belle collaboration, M.~Stari\v{c} {\em et~al.},
  \ifthenelse{\boolean{articletitles}}{\emph{{Search for a CP asymmetry in
  Cabibbo-suppressed $D^0$ decays}},
  }{}\href{https://doi.org/10.1016/j.physletb.2008.10.052}{Phys.\ Lett.\
  \textbf{B670} (2008) 190},
  \href{http://arxiv.org/abs/0807.0148}{{\normalfont\ttfamily
  arXiv:0807.0148}}\relax
\mciteBstWouldAddEndPuncttrue
\mciteSetBstMidEndSepPunct{\mcitedefaultmidpunct}
{\mcitedefaultendpunct}{\mcitedefaultseppunct}\relax
\EndOfBibitem
\bibitem{bib:cdfpaper}
CDF collaboration, T.~Aaltonen {\em et~al.},
  \ifthenelse{\boolean{articletitles}}{\emph{{Measurement of CP-violating
  asymmetries in $D^0\to\pi^+\pi^-$ and $D^0\to K^+K^-$ decays at CDF}},
  }{}\href{https://doi.org/10.1103/PhysRevD.85.012009}{Phys.\ Rev.\
  \textbf{D85} (2012) 012009},
  \href{http://arxiv.org/abs/1111.5023}{{\normalfont\ttfamily
  arXiv:1111.5023}}\relax
\mciteBstWouldAddEndPuncttrue
\mciteSetBstMidEndSepPunct{\mcitedefaultmidpunct}
{\mcitedefaultendpunct}{\mcitedefaultseppunct}\relax
\EndOfBibitem
\bibitem{CDF:2012qw}
CDF collaboration, T.~Aaltonen {\em et~al.},
  \ifthenelse{\boolean{articletitles}}{\emph{{Measurement of the difference of
  CP-violating asymmetries in $D^0 \to K^+K^-$ and $D^0 \to \pi^+\pi^-$ decays
  at CDF}}, }{}\href{https://doi.org/10.1103/PhysRevLett.109.111801}{Phys.\
  Rev.\ Lett.\  \textbf{109} (2012) 111801},
  \href{http://arxiv.org/abs/1207.2158}{{\normalfont\ttfamily
  arXiv:1207.2158}}\relax
\mciteBstWouldAddEndPuncttrue
\mciteSetBstMidEndSepPunct{\mcitedefaultmidpunct}
{\mcitedefaultendpunct}{\mcitedefaultseppunct}\relax
\EndOfBibitem
\bibitem{LHCB-PAPER-2011-023}
LHCb collaboration, R.~Aaij {\em et~al.},
  \ifthenelse{\boolean{articletitles}}{\emph{{Evidence for \CP violation in
  time-integrated $\Dz\to h^-h^+$ decay rates}},
  }{}\href{https://doi.org/10.1103/PhysRevLett.108.111602}{Phys.\ Rev.\ Lett.\
  \textbf{108} (2012) 111602},
  \href{http://arxiv.org/abs/1112.0938}{{\normalfont\ttfamily
  arXiv:1112.0938}}\relax
\mciteBstWouldAddEndPuncttrue
\mciteSetBstMidEndSepPunct{\mcitedefaultmidpunct}
{\mcitedefaultendpunct}{\mcitedefaultseppunct}\relax
\EndOfBibitem
\bibitem{LHCB-PAPER-2013-003}
LHCb collaboration, R.~Aaij {\em et~al.},
  \ifthenelse{\boolean{articletitles}}{\emph{{Search for direct \CP violation
  in $\Dz\to h^- h^+$ modes using semileptonic $\B$ decays}},
  }{}\href{https://doi.org/10.1016/j.physletb.2013.04.061}{Phys.\ Lett.\
  \textbf{B723} (2013) 33},
  \href{http://arxiv.org/abs/1303.2614}{{\normalfont\ttfamily
  arXiv:1303.2614}}\relax
\mciteBstWouldAddEndPuncttrue
\mciteSetBstMidEndSepPunct{\mcitedefaultmidpunct}
{\mcitedefaultendpunct}{\mcitedefaultseppunct}\relax
\EndOfBibitem
\bibitem{LHCB-PAPER-2014-013}
LHCb collaboration, R.~Aaij {\em et~al.},
  \ifthenelse{\boolean{articletitles}}{\emph{{Measurement of \CP asymmetry in
  $\Dz\to \Km\Kp$ and $\Dz\to \pim\pip$ decays}},
  }{}\href{https://doi.org/10.1007/JHEP07(2014)041}{JHEP \textbf{07} (2014)
  041}, \href{http://arxiv.org/abs/1405.2797}{{\normalfont\ttfamily
  arXiv:1405.2797}}\relax
\mciteBstWouldAddEndPuncttrue
\mciteSetBstMidEndSepPunct{\mcitedefaultmidpunct}
{\mcitedefaultendpunct}{\mcitedefaultseppunct}\relax
\EndOfBibitem
\bibitem{LHCB-PAPER-2015-055}
LHCb collaboration, R.~Aaij {\em et~al.},
  \ifthenelse{\boolean{articletitles}}{\emph{{Measurement of the difference of
  time-integrated \CP asymmetries in $\Dz\to \Km\Kp$ and $\Dz\to \pim\pip$
  decays}}, }{}\href{https://doi.org/10.1103/PhysRevLett.116.191601}{Phys.\
  Rev.\ Lett.\  \textbf{116} (2016) 191601},
  \href{http://arxiv.org/abs/1602.03160}{{\normalfont\ttfamily
  arXiv:1602.03160}}\relax
\mciteBstWouldAddEndPuncttrue
\mciteSetBstMidEndSepPunct{\mcitedefaultmidpunct}
{\mcitedefaultendpunct}{\mcitedefaultseppunct}\relax
\EndOfBibitem
\bibitem{LHCB-PAPER-2016-035}
LHCb collaboration, R.~Aaij {\em et~al.},
  \ifthenelse{\boolean{articletitles}}{\emph{{Measurement of \CP asymmetry in
  $\Dz\to \Kp\Km$ decays}},
  }{}\href{https://doi.org/10.1016/j.physletb.2017.01.061}{Phys.\ Lett.\
  \textbf{B767} (2017) 177},
  \href{http://arxiv.org/abs/1610.09476}{{\normalfont\ttfamily
  arXiv:1610.09476}}\relax
\mciteBstWouldAddEndPuncttrue
\mciteSetBstMidEndSepPunct{\mcitedefaultmidpunct}
{\mcitedefaultendpunct}{\mcitedefaultseppunct}\relax
\EndOfBibitem
\bibitem{Gersabeck:2011xj}
M.~Gersabeck {\em et~al.}, \ifthenelse{\boolean{articletitles}}{\emph{{On the
  interplay of direct and indirect CP violation in the charm sector}},
  }{}\href{https://doi.org/10.1088/0954-3899/39/4/045005}{J.\ Phys.\
  \textbf{G39} (2012) 045005},
  \href{http://arxiv.org/abs/1111.6515}{{\normalfont\ttfamily
  arXiv:1111.6515}}\relax
\mciteBstWouldAddEndPuncttrue
\mciteSetBstMidEndSepPunct{\mcitedefaultmidpunct}
{\mcitedefaultendpunct}{\mcitedefaultseppunct}\relax
\EndOfBibitem
\bibitem{LHCb-PAPER-2014-069}
LHCb collaboration, R.~Aaij {\em et~al.},
  \ifthenelse{\boolean{articletitles}}{\emph{{Measurement of indirect \CP
  asymmetries in $\Dz\to \Km\Kp$ and $\Dz\to \pim\pip$ decays using
  semileptonic $B$ decays}},
  }{}\href{https://doi.org/10.1007/JHEP04(2015)043}{JHEP \textbf{04} (2015)
  043}, \href{http://arxiv.org/abs/1501.06777}{{\normalfont\ttfamily
  arXiv:1501.06777}}\relax
\mciteBstWouldAddEndPuncttrue
\mciteSetBstMidEndSepPunct{\mcitedefaultmidpunct}
{\mcitedefaultendpunct}{\mcitedefaultseppunct}\relax
\EndOfBibitem
\bibitem{LHCb-PAPER-2016-063}
LHCb collaboration, R.~Aaij {\em et~al.},
  \ifthenelse{\boolean{articletitles}}{\emph{{Measurement of the \CP violation
  parameter $A_\Gamma$ in $\Dz\to \Kp\Km$ and $\Dz\to \pip\pim$ decays}},
  }{}\href{https://doi.org/10.1103/PhysRevLett.118.261803}{Phys.\ Rev.\ Lett.\
  \textbf{118} (2017) 261803},
  \href{http://arxiv.org/abs/1702.06490}{{\normalfont\ttfamily
  arXiv:1702.06490}}\relax
\mciteBstWouldAddEndPuncttrue
\mciteSetBstMidEndSepPunct{\mcitedefaultmidpunct}
{\mcitedefaultendpunct}{\mcitedefaultseppunct}\relax
\EndOfBibitem
\bibitem{bib:kagansokoloff}
A.~L. Kagan and M.~D. Sokoloff,
  \ifthenelse{\boolean{articletitles}}{\emph{{Indirect \CP violation and
  implications for \Dz--\Dzb and $B_s$--$\overline{B}_s$ mixing}},
  }{}\href{https://doi.org/10.1103/PhysRevD.80.076008}{Phys.\ Rev.\
  \textbf{D80} (2009) 076008},
  \href{http://arxiv.org/abs/0907.3917}{{\normalfont\ttfamily
  arXiv:0907.3917}}\relax
\mciteBstWouldAddEndPuncttrue
\mciteSetBstMidEndSepPunct{\mcitedefaultmidpunct}
{\mcitedefaultendpunct}{\mcitedefaultseppunct}\relax
\EndOfBibitem
\bibitem{Du:2006jc}
D.-S. Du, \ifthenelse{\boolean{articletitles}}{\emph{{CP violation for neutral
  charmed meson decays to CP eigenstates}},
  }{}\href{https://doi.org/10.1140/epjc/s10052-007-0242-6}{Eur.\ Phys.\ J.\
  \textbf{C50} (2007) 579},
  \href{http://arxiv.org/abs/hep-ph/0608313}{{\normalfont\ttfamily
  arXiv:hep-ph/0608313}}\relax
\mciteBstWouldAddEndPuncttrue
\mciteSetBstMidEndSepPunct{\mcitedefaultmidpunct}
{\mcitedefaultendpunct}{\mcitedefaultseppunct}\relax
\EndOfBibitem
\bibitem{LHCb-PAPER-2016-062}
LHCb collaboration, R.~Aaij {\em et~al.},
  \ifthenelse{\boolean{articletitles}}{\emph{{Measurement of \Bd, \Bs, \Bp and
  \Lb production asymmetries in $7$ and $8$\tev\ proton-proton collisions}},
  }{}\href{https://doi.org/10.1016/j.physletb.2017.09.023}{Phys.\ Lett.\
  \textbf{B774} (2017) 139},
  \href{http://arxiv.org/abs/1703.08464}{{\normalfont\ttfamily
  arXiv:1703.08464}}\relax
\mciteBstWouldAddEndPuncttrue
\mciteSetBstMidEndSepPunct{\mcitedefaultmidpunct}
{\mcitedefaultendpunct}{\mcitedefaultseppunct}\relax
\EndOfBibitem
\bibitem{LHCb-PAPER-2013-033}
LHCb collaboration, R.~Aaij {\em et~al.},
  \ifthenelse{\boolean{articletitles}}{\emph{{Measurement of the
  flavour-specific \CP-violating asymmetry $a_{\mathrm{sl}}^s$ in $\Bs$
  decays}}, }{}\href{https://doi.org/10.1016/j.physletb.2013.12.030}{Phys.\
  Lett.\  \textbf{B728} (2014) 607},
  \href{http://arxiv.org/abs/1308.1048}{{\normalfont\ttfamily
  arXiv:1308.1048}}\relax
\mciteBstWouldAddEndPuncttrue
\mciteSetBstMidEndSepPunct{\mcitedefaultmidpunct}
{\mcitedefaultendpunct}{\mcitedefaultseppunct}\relax
\EndOfBibitem
\bibitem{LHCb-PAPER-2012-026}
LHCb collaboration, R.~Aaij {\em et~al.},
  \ifthenelse{\boolean{articletitles}}{\emph{{Measurement of the $\Dpm$
  production asymmetry in $7$\tev $\proton\proton$ collisions}},
  }{}\href{https://doi.org/10.1016/j.physletb.2012.11.038}{Phys.\ Lett.\
  \textbf{B718} (2013) 902},
  \href{http://arxiv.org/abs/1210.4112}{{\normalfont\ttfamily
  arXiv:1210.4112}}\relax
\mciteBstWouldAddEndPuncttrue
\mciteSetBstMidEndSepPunct{\mcitedefaultmidpunct}
{\mcitedefaultendpunct}{\mcitedefaultseppunct}\relax
\EndOfBibitem
\bibitem{LHCb-PAPER-2012-009}
LHCb collaboration, R.~Aaij {\em et~al.},
  \ifthenelse{\boolean{articletitles}}{\emph{{Measurement of the $\Dsp$--$\Dsm$
  production asymmetry in $7$\tev $\proton\proton$ collisions}},
  }{}\href{https://doi.org/10.1016/j.physletb.2012.06.001}{Phys.\ Lett.\
  \textbf{B713} (2012) 186},
  \href{http://arxiv.org/abs/1205.0897}{{\normalfont\ttfamily
  arXiv:1205.0897}}\relax
\mciteBstWouldAddEndPuncttrue
\mciteSetBstMidEndSepPunct{\mcitedefaultmidpunct}
{\mcitedefaultendpunct}{\mcitedefaultseppunct}\relax
\EndOfBibitem
\bibitem{Alves:2008zz}
LHCb collaboration, A.~A. Alves~Jr.\ {\em et~al.},
  \ifthenelse{\boolean{articletitles}}{\emph{{The \lhcb detector at the LHC}},
  }{}\href{https://doi.org/10.1088/1748-0221/3/08/S08005}{JINST \textbf{3}
  (2008) S08005}\relax
\mciteBstWouldAddEndPuncttrue
\mciteSetBstMidEndSepPunct{\mcitedefaultmidpunct}
{\mcitedefaultendpunct}{\mcitedefaultseppunct}\relax
\EndOfBibitem
\bibitem{LHCb-DP-2014-002}
LHCb collaboration, R.~Aaij {\em et~al.},
  \ifthenelse{\boolean{articletitles}}{\emph{{LHCb detector performance}},
  }{}\href{https://doi.org/10.1142/S0217751X15300227}{Int.\ J.\ Mod.\ Phys.\
  \textbf{A30} (2015) 1530022},
  \href{http://arxiv.org/abs/1412.6352}{{\normalfont\ttfamily
  arXiv:1412.6352}}\relax
\mciteBstWouldAddEndPuncttrue
\mciteSetBstMidEndSepPunct{\mcitedefaultmidpunct}
{\mcitedefaultendpunct}{\mcitedefaultseppunct}\relax
\EndOfBibitem
\bibitem{Kodama:1991ij}
E653 collaboration, K.~Kodama {\em et~al.},
  \ifthenelse{\boolean{articletitles}}{\emph{{Measurement of the relative
  branching fraction $\Gamma (D^0 \to K \mu \nu) / \Gamma (D^0 \to \mu X)$}},
  }{}\href{https://doi.org/10.1103/PhysRevLett.66.1819}{Phys.\ Rev.\ Lett.\
  \textbf{66} (1991) 1819}\relax
\mciteBstWouldAddEndPuncttrue
\mciteSetBstMidEndSepPunct{\mcitedefaultmidpunct}
{\mcitedefaultendpunct}{\mcitedefaultseppunct}\relax
\EndOfBibitem
\bibitem{Hulsbergen2005566}
W.~D. Hulsbergen, \ifthenelse{\boolean{articletitles}}{\emph{{Decay chain
  fitting with a Kalman filter}},
  }{}\href{https://doi.org/10.1016/j.nima.2005.06.078}{Nucl.\ Instrum.\ Meth.\
  \textbf{A552} (2005) 566},
  \href{http://arxiv.org/abs/physics/0503191}{{\normalfont\ttfamily
  arXiv:physics/0503191}}\relax
\mciteBstWouldAddEndPuncttrue
\mciteSetBstMidEndSepPunct{\mcitedefaultmidpunct}
{\mcitedefaultendpunct}{\mcitedefaultseppunct}\relax
\EndOfBibitem
\bibitem{Johnson:1949zj}
N.~L. Johnson, \ifthenelse{\boolean{articletitles}}{\emph{{Systems of frequency
  curves generated by methods of translation}},
  }{}\href{https://doi.org/10.1093/biomet/36.1-2.149}{Biometrika \textbf{36}
  (1949) 149}\relax
\mciteBstWouldAddEndPuncttrue
\mciteSetBstMidEndSepPunct{\mcitedefaultmidpunct}
{\mcitedefaultendpunct}{\mcitedefaultseppunct}\relax
\EndOfBibitem
\bibitem{LHCb-PAPER-2019-006}
LHCb collaboration, R.~Aaij {\em et~al.},
  \ifthenelse{\boolean{articletitles}}{\emph{{Observation of \CP violation in
  charm decays}},
  }{}\href{http://arxiv.org/abs/1903.08726}{{\normalfont\ttfamily
  arXiv:1903.08726}}, {accepted by Phys. Rev. Lett.}\relax
\mciteBstWouldAddEndPunctfalse
\mciteSetBstMidEndSepPunct{\mcitedefaultmidpunct}
{}{\mcitedefaultseppunct}\relax
\EndOfBibitem
\bibitem{HFLAV16}
Heavy Flavor Averaging Group, Y.~Amhis {\em et~al.},
  \ifthenelse{\boolean{articletitles}}{\emph{{Averages of $b$-hadron,
  $c$-hadron, and $\tau$-lepton properties as of summer 2016}},
  }{}\href{https://doi.org/10.1140/epjc/s10052-017-5058-4}{Eur.\ Phys.\ J.\
  \textbf{C77} (2017) 895},
  \href{http://arxiv.org/abs/1612.07233}{{\normalfont\ttfamily
  arXiv:1612.07233}}, {updated results and plots available at
  \href{https://hflav.web.cern.ch}{{\texttt{https://hflav.web.cern.ch}}}}\relax
\mciteBstWouldAddEndPuncttrue
\mciteSetBstMidEndSepPunct{\mcitedefaultmidpunct}
{\mcitedefaultendpunct}{\mcitedefaultseppunct}\relax
\EndOfBibitem
\bibitem{PDG2018}
Particle Data Group, M.~Tanabashi {\em et~al.},
  \ifthenelse{\boolean{articletitles}}{\emph{{\href{http://pdg.lbl.gov/}{Review
  of particle physics}}},
  }{}\href{https://doi.org/10.1103/PhysRevD.98.030001}{Phys.\ Rev.\
  \textbf{D98} (2018) 030001}\relax
\mciteBstWouldAddEndPuncttrue
\mciteSetBstMidEndSepPunct{\mcitedefaultmidpunct}
{\mcitedefaultendpunct}{\mcitedefaultseppunct}\relax
\EndOfBibitem
\end{mcitethebibliography}

\end{document}